\documentclass{appolb}
\usepackage{epsfig}
\newcommand{\Li}[2]{{\mbox{Li}}_{#1}\left(#2\right)}    
\newcommand{\ep}{\varepsilon}
\begin{document}
\title{ $O(\alpha \alpha_s)$ relation between pole- and $\overline{\rm
MS}$-mass of the t-quark\thanks{Based on the talk given by M.Yu.K. at the 
XXVII International Conference of Theoretical Physics, Ustro\'n, Poland,
September 15-21, 2003.\\ Work supported in part by DFG under Contract
SFB/TR~9-03 and by the European Community's Human Potential Programme
under Contract HPRN-CT-2002-00311, EURIDICE .  }
%
}
\author{F.~Jegerlehner and M.~Yu.~Kalmykov
\thanks{On leave from BLTP, JINR, 141980 Dubna, Russia.}
\address{DESY Zeuthen, Platanenallee 6, D-15738, Zeuthen, Germany}
}
\maketitle
\begin{abstract}
The $O(\alpha \alpha_s)$ contribution to the relationship between the
$\overline{\rm MS}$- and the pole-mass of the t-quark propagator within
the Standard Model is reviewed.  At the same order also the corrections
to the top-Yukawa coupling is discussed.  We furthermore present the
exact analytic expression for the gaugeless limit.
\end{abstract}
  
\vspace*{-10.5cm}
\begin{flushright}
{DESY 03-180  }\\
{SFB/CPP-03-50}\\
{hep-ph/0310361} \\
{October 2003}
\end{flushright}
\vspace*{8cm}    
\section{Introduction}
The Standard Model (SM) belongs to the class of {\it
renormalizable} quantum field theories \cite{SM} which means, in
particular, that a restricted number of input parameters suffice for
theoretical predictions of any process.  The concrete choice of input
parameters defines a specific {\it renormalization scheme}.  The given
set of independent parameters has to be extracted from an appropriate
set of experimentally measured quantities.  If we would be able to
perform perturbative calculations to all orders all renormalization
schemes would be equivalent.  However, in practice, only the first few
coefficients are known, so that predictions depend on the choice of
the scheme.  Such dependence on the truncation of the perturbative
series is known as {\it scheme dependence}.  In general, the
difference between two schemes is of next higher order in the
perturbation expansion.  For higher order calculations those schemes
are preferable for which the uncalculated higher order corrections are
small.  Of course, to find such a preferred scheme requires to perform
calculations in different schemes\cite{scheme}.  Another possibility
is to find the scheme transition relations by calculating the input
parameters in one scheme in terms of the input parameters of another
scheme order by order in perturbation theory.  For electroweak
calculations a natural and generally accepted scheme is the so called {\it
on-shell scheme}~\cite{Sirlin80}--\cite{Denner93}, where, in addition
to the fine structure constant (and/or the Fermi constant), the pole
masses of particles serve as input parameters. However, for quarks the
pole mass suffers from renormalon contributions\cite{renormalon} which
affect seriously the convergence of the perturbation expansion. This
is one of the main reasons why for quarks the $\overline{\rm MS}$-mass
appears to be a better input parameter. A good illustration of this
point is the behavior of the QCD corrections to the $\rho$-parameter,
which are large when $\Delta \rho$ is expressed in terms of the
pole-mass. In contrast, in terms of the $\overline{\rm MS}$-mass the
expansion coefficients are much smaller\cite{rho,MS:ON}.

The relation between pole- and $\overline{\rm MS}$-mass of quarks has
been calculated including one-, two- and leading three-loop
corrections. The one--loop results at $O(\alpha_s)$ and $O(\alpha)$ have
been presented in~\cite{Tarrach} and e.g. in~\cite{BHS86}\footnote{(see
also Eq.~(B.5) in Appendix B of~\cite{poleII})}, respectively. The
two--loop $O(\alpha_s^2)$ correction is given in~\cite{Broadhurst}, and
the same result was obtained via regularization by dimensional reduction
in~\cite{red}.  The renormalized off-shell fermion propagator of order
$O(\alpha_s^2)$ has been worked out in~\cite{FJTV}. Only recently,
in~\cite{pole3}, the three--loop $O(\alpha_s^3)$ correction has been
published. Finally, the two--loop $O(\alpha \alpha_s)$ and $O(\alpha^2)$
corrections have been calculated in the approximation of vanishing
electroweak gauge couplings~\cite{rho3}. Our recent
calculation~\cite{our}, extends previous two--loop $O(\alpha
\alpha_s)$ calculations of the gauge boson self--energies~\cite{QCD}
and the SM $O(\alpha^2)$ corrections to the relation between the pole-
and the $\overline{\rm MS}$-mass of the gauge bosons $Z$ and $W$,
presented in~\cite{poleI,poleII}.

\section{Definitions}
The definition of the top-quark pole mass has been discussed
in~\cite{toppole}. In general, the position of the pole of a fermion
propagator, which defines the pole-mass, is given by the formal
solution for the momentum $\hat{p}=i\tilde{M}$, at which the inverse of
the connected full propagator equal zero
\begin{equation}
i \hat{p} + m - \tilde{\Sigma}(p, m, \ldots)= 0 \;. 
\label{pole}
\end{equation}
The ``mass'' $\tilde{M}$ is a complex number, i.e., $\tilde{M} \equiv
M'-\:\frac{i}{2}\: \Gamma'$. The latter parameters are related to the
pole mass $M$ and the on-shell width $\Gamma$, which are parametrizing
the pole of the squared transition matrix element $|T|^2$ analogous to the
boson case, by (see
~\cite{our})
\begin{equation}
\tilde{M}^2=M^2-iM\Gamma=M^{'2}-\Gamma^{'2}/4-iM'\Gamma' \;, 
\end{equation}
such that
\begin{equation}
M=\sqrt{M^{'2}-\Gamma^{'2}/4} \;, \quad \Gamma=\frac{M'}{M}\Gamma' \;. 
\label{properMG}
\end{equation}
Since $M=M' + O(\alpha^2)$ and $\Gamma=\Gamma' + O(\alpha^2)$ for the
$O(\alpha \alpha_s)$ terms considered in this paper we can identify
$M=M'$ and $\Gamma=\Gamma'$ in the following.

For the remainder of the paper we will adopt the following notation:
capital $M \simeq {\rm Re}\:\tilde{M}$ always denotes the pole mass;
lower case $m$ stands for the renormalized mass in the 
$\overline{\rm MS}$ scheme, while $m_0$ denotes the bare mass. 
The on--shell width is
given by $\Gamma \simeq -2\: {\rm Im}\:\tilde{M}$. In addition we use
$e$, $g$ and $g_s$ to denote the $U(1)_{\rm em}$, $SU(2)_{\rm L}$ and
$SU(3)_{\rm c}$ couplings of the SM in the $\overline{\rm MS}$  scheme.

In perturbation theory (\ref{pole}) is to be solved order by order.
For this aim we expand the self--energy function about the lowest
order solution $\hat{p}=i m_0$:
$$
\tilde{\Sigma}(p, m, \ldots) = \tilde{\Sigma} \left. \right|_{\hat{p}=i m_0}
\!+\! \left(i \hat{p} \!+\! m_0  \right) \left. \left[\tilde{\Sigma}^{'}\right]
\right|_{\hat{p}=i m_0}
\!+\! \cdots
= \Sigma \!+\! \left(i \hat{p} \!+\! m_0  \right) \Sigma' 
\!+\! \cdots
$$
To two loops we then have the solution 
\begin{eqnarray}
\hspace{-1cm}
\frac{\tilde{M}}{m} = 1 + \Sigma_1 + \Sigma_2 + \Sigma_1 \Sigma^{'}_1 \;, 
\label{polemass}
\end{eqnarray} %
where $\Sigma_L$ is the bare ($m=m_0$) or  $\overline{\rm MS}$--renormalized 
($m$ the  $\overline{\rm MS}$--mass) $L$-loop contribution to the amplitude. 
According to Eq.~(\ref{polemass}) we need to calculate propagator-type
diagrams up to two loops on--shell.  In order to get manifestly gauge
invariant results the Higgs tadpole diagrams must be
included~\cite{FJ81}.  As we have elaborated in~\cite{poleII} the
inclusion of the tadpoles is mandatory also for the self--consistency of
the renormalization group (RG).

For our calculation all diagrams have been generated with the help of
${\bf QGRAF}$~\cite{qgraf}. The C-program {\bf DIANA}~\cite{diana} then
was used together with the set of Feynman rules extracted from the
package {\bf TLAMM}~\cite{tlamm} to produce the FORM input which is
suitable for the package {\bf ONSHELL2}~\cite{onshell2} and/or for
another package based on Tarasov's recurrence relations~\cite{T97a}.
The relevant master-integrals have been calculated
analytically\footnote{Of course, another possibility is to apply
directly numerical programs, some of which are discussed
in~\cite{programms}.} with the help of techniques developed recently
in~\cite{DK}.
\section{The gaugeless limit}

The complete $O(\alpha \alpha_s)$ result of the calculation within the
SM is given in our recent publication~\cite{our}. Here we present some
details concerning the renormalization and present the complete ${\cal
O}(\alpha \alpha_s)$ answer for the so called ``gaugeless limit''
approximation of the SM. This limit corresponds to the case $M_H, m_t
\gg M_W$ and can be deduced from the Lagrangian of the SM in the
approximation $g,g' \to 0, v^2 \neq 0$. It allows us a simplified
calculation of the leading top-quark mass corrections to physical
observables, like the $\rho$-parameter or corrections to $Z\to b
\overline {b}$ decay~\cite{rho,rho3}.

The mass renormalization constant $Z_t$ in the $\overline{\rm MS}$ scheme at 
two loops may be written in the form
\begin{eqnarray}
m_{t,0} & = &  m_t(\mu^2)\;Z_t
= m_t(\mu^2)\:
\Biggl( 1
+ \frac{g^2(\mu^2)}{16\pi^2} \frac{m_t^2}{m_W^2}\;\frac{1}{\ep} Z_\alpha^{(1,1)}
+ \frac{\alpha_s(\mu^2)}{4 \pi}\;\frac{1}{\ep} Z_{\alpha_s}^{(1)}
\nonumber \\ && \hspace{5mm}
+ \frac{\alpha_s(\mu^2)}{4 \pi } \frac{g^2(\mu^2)}{16\pi^2} \frac{m_t^2}{m_W^2}
\left( \frac{1}{\ep} Z_{\alpha \alpha_s}^{(2,1)} +
\frac{1}{\ep^2} Z_{\alpha \alpha_s}^{(2,2)}
\right)
+ {\cal O}(g^4, \alpha_s^2)
\Biggr),
\label{baremsb}
\end{eqnarray}
where $\alpha_s = g_s^2/4 \pi$ 
and 
\begin{eqnarray}
Z_\alpha^{(1,1)} & = &
- \frac{3}{8} \frac{m_H^2}{m_t^2}
+ \frac{3}{8} 
+ N_c \frac{m_t^2}{m_H^2} \;, \quad 
Z_{\alpha_s}^{(1)}  = - 3 C_f.
\label{RG:one-loop}
\end{eqnarray}
In our calculation we obtained the two-loop renormalization constants
$Z_{\alpha \alpha_s}^{(2,1)}$ and $Z_{\alpha \alpha_s}^{(2,2)}$
\begin{eqnarray}
&&
Z_{\alpha \alpha_s}^{(2,2)} = C_f \Biggl[
- 9 N_c \frac{m_t^2}{m_H^2}
+ \frac{9}{8} \frac{m_H^2}{m_t^2}
- \frac{9}{4} 
\Biggr] \;, \quad 
Z_{\alpha \alpha_s}^{(2,1)} =
 C_f \Biggl[
2 N_c \frac{m_t^2}{m_H^2}
+ \frac{3}{2} 
\Biggr] \;,
\label{Z2}
\end{eqnarray}
where, in the SM, $C_f = 4/3$, $N_c = 3$.  We may use the SM
renormalization group equations to cross-check the $1/\ep^2$-- and
$1/\ep$--terms \cite{poleII,RG}.  The coefficient $Z_{\alpha
\alpha_s}^{(2,2)}$ may be calculated from the RG relations which allow
one to predict the leading higher order poles in terms of the RG
coefficients (see Eq.(4.39) in~\cite{our} ).  The terms proportional to
$1/\ep$ may be deduced from the RG equations calculated in the unbroken
phase.  It has been shown~\cite{poleII,RG} that in the $\overline{\rm
MS}$ scheme we may write
\begin{equation}
m_t^2 (\mu^2) = \frac{1}{2} \frac{Y^2_t(\mu^2)}{\lambda(\mu^2)}\;
m^2(\mu^2) \;,
\end{equation}
where $m^2$ and $\lambda$ are the parameters of the symmetric scalar
potential $V$
$$
V = \frac{\mu^2}{2} \phi^2 + \frac{\lambda}{24} \phi^4
$$
and $Y_t$ is the top-quark Yukawa coupling. As a consequence
we get the following relation for the anomalous dimension $\gamma_t$ 
of the mass of the top-quark 
\begin{eqnarray}
\gamma_t & = & \gamma_{Y} + \frac{1}{2} \gamma_{m^2} - \frac{1}{2} \frac{\beta_\lambda}{\lambda} \;,
\label{SM<->F}
\end{eqnarray}
where the relevant RG results in the gaugeless limit up to ${\cal O} (g^4)$ are 
\cite{RG2}
\begin{eqnarray}
&&
\gamma_{m^2} \equiv
\frac{1}{m^2} \mu^2 \frac{d}{d \mu^2} m^2  =
\frac{1}{16 \pi^2}
\Biggl[ \lambda + 3 Y_t^2  \Biggr] 
+ 20  \frac{g_s^2 Y_t^2}{(16 \pi^2)^2}
\; ,
\nonumber \\
&&
\beta_\lambda \equiv \mu^2 \frac{d}{d \mu^2} \lambda
= \frac{1}{(16 \pi^2)}
\Biggl[
2 \lambda^2
+ 6 \lambda Y_t^2
- 18 Y_t^4
\Biggr]
+ \frac{g_s^2 Y_t^2 }{(16 \pi^2)^2}
\Biggl[
40 \lambda - 96 Y_t^2
\Biggr]
\;,
\nonumber \\
&&
\gamma_{Y} \equiv
\frac{1}{Y_t} \mu^2 \frac{d}{d \mu^2} Y_t  =
\frac{1}{16 \pi^2}
\Biggl[ \frac{9}{4} Y_t^2 - 4 g_s^2  \Biggr]
+
18 \frac{g_s^2 Y_t^2}{(16 \pi^2)^2} 
\;.
\label{RG}
\end{eqnarray}
At the same time, the anomalous dimension $\gamma_t$ can be related with 
the renormalization constant (see~\cite{poleI} for details). 
In our case we get 
$$
\gamma_t = \frac{m_t^2}{m_W^2} \Biggl[
\frac{g^2}{16 \pi^2} Z_\alpha^{(1,1)} 
+ \frac{g^2 g_s^2}{ (16 \pi^2)^2} 2 Z_{\alpha \alpha_s}^{(2,1)}
\Biggr] 
+ \frac{g_s^2}{16 \pi^2} Z_{\alpha_s}^{(1)}
\;. 
$$
Finally, the parameter relations
$
Y_t^2 = \frac{2 m_t^2}{v^2} \;, \quad
\lambda = \frac{3 m_H^2}{v^2} \;, \quad
g^2 = \frac{4 m_W^2}{v^2} \;, 
$
provide the bridge between Eqs.~(\ref{SM<->F}) and
(\ref{RG}) and our Eqs.~(\ref{RG:one-loop}) and (\ref{Z2}).

After performing the UV renormalization the MS renormalized amplitudes 
are finite.  The relation between the top--propagator pole $\tilde{M}$ 
and the $\overline{\rm MS}$ mass $m_t$ can be written as
\begin{equation}
\frac{\tilde{M}}{m_t} = 1 
+ \Sigma_{1,{\overline{\rm MS}}}
+ \Biggl \{ \Sigma_2 + \Sigma_1 \Sigma_1^{'}
\Biggr\}_{\overline{\rm MS}}
+ {\cal O} (g^4,\alpha_s^2)
\;,
\label{result}
\end{equation}
where 
\begin{eqnarray}
&& 
\Sigma_{1,{\overline{\rm MS}}}
= 
\frac{\alpha_s}{4 \pi } C_f \Biggl[ 4 - 3 \ln \frac{m_t^2}{\mu^2}\Biggr]  
+ 
\frac{g^2}{16 \pi^2} \frac{m_t^2}{m_W^2}
\Biggl[
Z_{\alpha}^{(1,1)} \ln \frac{m_t^2}{\mu^2}
+ \frac{(1+y^2)}{2 y} 
\nonumber \\  && 
- N_c \frac{m_t^2}{m_H^2}
- \frac{m_t^4}{8 m_H^4} \ln (1+y)  
+ \frac{m_t^2}{8 m_H^2} \frac{(3+y^2)}{(1+y)} \ln y 
- i \pi  \frac{1}{8} 
 \Biggr], \; 
\\&& 
\Biggl \{ \Sigma_2 + \Sigma_1 \Sigma_1^{'} \Biggr\}_{\overline{\rm MS}}
= C_f \frac{\alpha_s}{4 \pi} \frac{g^2}{16\pi^2} \frac{m_t^2}{m_W^2}
\Biggl(
\ln^2 \frac{m_t^2}{\mu^2}
\Biggl[
    \frac{9}{8} \frac{m_H^2}{m_t^2}
- 9 N_c \frac{m_t^2}{m_H^2}
- \frac{9}{4} 
\Biggr]
\nonumber \\  && 
+  \ln  \frac{m_t^2}{\mu^2} 
\Biggl[
9 
\!+\! 11 N_c \frac{m_t^2}{m_H^2}
\!-\! \frac{3 m_t^4}{8 m_H^4} \ln (1\!+\!y) 
\!+\! \frac{3 m_t^2}{8 m_H^2} \frac{(y^2 \!+\! 6y \!-\! 3)}{(1\!+\!y)} \ln y 
\!+\! i \pi \frac{9}{8} 
\Biggr]
\nonumber \\  && 
+ \zeta_2
\Biggl\{
 \frac{3}{2  y}
+ \frac{9}{2} y
+ \frac{3}{4} y^2
\Biggr\}
- \frac{y}{(1+y)^2}
\Biggl\{
  \frac{11 (1+y^2) (1+y)^2 }{8 y^2}
+ 8 N_c
\Biggr\}
\nonumber \\ && 
+ \frac{(1-y)^2}{y^2} \ln y
\Biggl[ \ln (1-y) + \frac{1}{2} \ln (1+y) \Biggr]
\Biggl[ (1-y^2) - \frac{1}{2} (1+y^2) \ln y \Biggr]
\nonumber \\ &&
- \frac{1}{8} \frac{2+8 y - 10 y^2 - 3y^3}{y } \ln^2 y
+ \frac{1}{8} \frac{(1+y )(6 - 63 y + 5 y^2)}{y } \ln y
\nonumber \\ &&
- \frac{1}{8} \frac{(1+y)^2 (5 - 62 y + 5 y^2)}{y^2} \ln (1 + y)
- \frac{3}{2} \zeta_2 \ln (1+y) \frac{(1-y)^2 (1+y^2)}{y^2}
\nonumber \\ && 
+ \frac{(1-y) (1+y) }{y^2}
\Biggl\{
\frac{(5 - 28 y + 5 y^2)}{4} \Li{2}{-y} + (1-y)^2 \Li{2}{y}
\Biggr\}
\nonumber \\ && 
+ \frac{(1-y)^2 (1+y^2)}{y^2}
\Biggl\{
\frac{3}{2} \Biggl[2 \Li{3}{ y}  + \Li{3}{-y}  \Biggr]
- \ln y \Biggl[ 2 \Li{2}{y} + \Li{2}{-y}  \Biggr]
\Biggr\} 
\nonumber \\ &&
- \frac{259}{16}
- \frac{15}{4} \zeta_2 
- \frac{3}{2} \zeta_3
- i \pi \left[ 
\frac{17}{8} - \zeta_2
\right]
\Biggr)
\; ,
\label{MS2:subtracted}
\end{eqnarray}
where
$$
y = \frac{1-\sqrt{1-\frac{4 m_t^2}{m_H^2}}} {1+\sqrt{1-\frac{4 m_t^2}{m_H^2}}} 
\equiv 
\frac{1-\sqrt{1-\frac{6 Y_t^2}{\lambda}}} {1+\sqrt{1-\frac{6 Y_t^2}{\lambda}}} \; .
$$ As the top is an unstable particle the pole of the propagator has an
imaginary part which is related up to a sign to the width $\Gamma_t$ divided
by two.  In the gaugeless limit approximation it is equal to
\begin{eqnarray}
\frac{\Gamma_t}{M_t} = 
\frac{\alpha}{2 \sin^2 \theta_W^{OS}} \frac{1}{8}  \frac{M_t^2}{M_W^2}
\Biggl[1 + \frac{\alpha_s}{4 \pi } C_f \left( 5  - 8 \zeta_2 \right)
\Biggr] \;. 
\end{eqnarray}
Very often the inverse of the relation (\ref{result}) is required.  To
that end we have to solve the real part of (\ref{polemass})
iteratively for $m_t$ and to express all $\overline{\rm MS}$
parameters in terms of on-shell ones. 
The ${\cal O }(\alpha \alpha_s)$ solution to two loops reads 

\begin{eqnarray}
&& \hspace{-7mm}
\frac{m_t}{M_t }
\!=\! 
1 
\!-\! {\rm Re} \Sigma_{1,\overline{\rm MS}}
\!-\! {\rm Re} \Biggl \{ \Sigma_2 \!+\! \Sigma_1 \Sigma_1^{'}
\Biggr\}_{\overline{\rm MS}}
\!+\! \frac{2 \alpha_s \alpha }{(4 \pi)^2 \sin^2 \theta_W^{OS}}
\Biggl\{
Z_{\alpha_s}^{(1)}~{\rm Re}~\Sigma_{1,\overline{\rm MS}}^\alpha
\nonumber\\ && \hspace{-7mm}
+  
\Sigma_{1,\overline{\rm MS}}^{\alpha_s} 
\left( 
\frac{M_t^2}{M_W^2} 
\Biggl[ 
\Delta X_\alpha^{(1)} 
\!+\! Z_\alpha^{(1,1)} 
\!+\! \ln \frac{M_t^2}{\mu^2} \left(\frac{3}{8} \!+\! 2 N_c \frac{M_t^2}{M_H^2} \right) 
\Biggr]
\!+\! {\rm Re} \Sigma_{1,\overline{\rm MS}}^\alpha 
\right)
\Biggr\},
\label{reverse:a}
\end{eqnarray}
where $\Sigma_{1,\overline{\rm MS}}^j$ ($j = \alpha, \alpha_s$) means
that only the $''j''$ part of the one-loop MS renormalized amplitude is
to be taken into account and $\Delta X^{(1)}_{\alpha}$ denotes the real
part of the derivative of the one-loop amplitude with respect to the
top-mass: $$
\Delta  X^{(1)}_{\alpha}  
= \frac{M_H^4}{8 M_t^4}  \ln (1+Y) 
- 2 N_c \frac{M_t^2}{M_H^2}
- \frac{(1\!+\!Y) (3\!+\!Y)}{8} \ln Y 
+ \frac{(1\!-\!Y)^2}{4 Y} \;, 
$$
with 
$
Y = \frac{1-\sqrt{1-\frac{4 M_t^2}{M_H^2}}} {1+\sqrt{1-\frac{4 M_t^2}{M_H^2}}} \;. 
$
It is interesting to compare the result (\ref{reverse:a}) calculated at 
$\mu=M_t$ with a similar relation calculated in the full SM 
(see Eq.(5.57) in~\cite{our}). 
The difference can be written in the following form 
\begin{equation}
\frac{m_t^{SM}(M_t) \!-\! m_t^{GL}(M_t)}{M_t}
\!=\! 
\frac{\alpha}{4 \pi \sin^2 \theta_W^{OS}} \frac{M_t^2}{M_W^2} 
\Biggl[ 
a \frac{M_t^2}{M_H^2} \left( 1 \!-\! 4 \frac{\alpha_s}{4 \pi} C_f \right) 
\!+\! b 
\!+\! c \frac{\alpha_s}{4 \pi} C_f 
\Biggr]
\end{equation}
where the constants $a,b,c$ depend only on the values of the masses of 
the gauge bosons $W,Z$ and the top-quark. 
Numerically, taking the input parameter values 
$M_W$ = 80.419 GeV, $M_Z$ = 91.188 GeV and $M_t$ = 174.3 GeV,
we obtain
\begin{eqnarray}
&& 
a = 
- \frac{1}{2} \frac{M_W^4}{M_t^4} \left( 1 - 3 \ln \frac{M_W^2}{M_t^2} \right)
- \frac{1}{4} \frac{M_Z^4}{M_t^4} \left( 1 - 3 \ln \frac{M_Z^2}{M_t^2} \right)
\sim  -0.21934
\;, 
\nonumber \\ &&
b  =  -0.07978 \;, \quad c  =  -0.429164 \;.
\end{eqnarray}

\section{$O(\alpha \alpha_s)$ correction to the top-Yukawa coupling}

In general, the concept of a quark mass is convention dependent.  In
electroweak theory we have the possibility to consider instead of the
mass of the top-quark, for example, the top-Yukawa coupling.  The
one-loop electroweak corrections ${\cal O}(\alpha)$ to the relation
between the Yukawa coupling and the pole parameters has been calculated
first in~\cite{Yukawa}.  Our result~\cite{our} allows us to extract the
${\cal O}(\alpha \alpha_s)$ correction to the top-Yukawa coupling. The
starting point is the relation between the Fermi constant $G_F$ and the
pole parameters of the SM, which may be written in the following
form~\cite{Sirlin80}: $$
\sin^2 \theta_W M_W^2 (1-\overline{\Delta} r ) = \frac{\pi \alpha(M_Z)}{ \sqrt{2} G_F} \;, 
\quad \overline{\Delta} r \equiv \Delta r - \Delta \alpha \;, 
$$
where for $\Delta r $ we use parametrization proposed in \cite{CHJ}.
Using the transition from the on-shell to the $\overline{\rm MS}$ parameters
of the SM (see~\cite{RG} for details) we get the following expression 
for
$v_{MS}^2(\mu^2) \equiv 1/\sqrt{2} G_F(\mu^2)$:
\begin{eqnarray}
v_{MS}^2(\mu^2) = \frac{1}{ \sqrt{2} G_F} \frac{1}{1-\overline{\Delta}r}
\Biggl[\frac{m_W^2 (\mu^2)}{M_W^2}\Biggr]  
\Biggl[\frac{\alpha(M_Z)}{\alpha_{MS}(\mu^2)} \Biggr]  
\Biggl[\frac{\sin^2 \theta^{MS}_W(\mu^2)}{\sin^2 \theta_W^{OS}}\Biggr]  \; , 
\end{eqnarray}
such that we have 

\begin{eqnarray}
\frac{y_t(\mu^2)}{M_t 2^{3/4}G_F^{1/2}}
= \frac{m_t(\mu^2)}{M_t}
\sqrt{(1-\overline{\Delta}r) 
\frac{\alpha_{MS}(\mu^2)}{\alpha(M_Z)} 
\frac{M_W^2}{m_W^2 (\mu^2)} 
\frac{\sin^2 \theta_W^{OS}}{\sin^2 \theta^{MS}_W(\mu^2)}} \;. 
\end{eqnarray}
This is our basic expression. Expanding each relation in powers of the
coupling constants $\alpha$ and $\alpha_s$ the correction to the top-Yukawa
coupling can be extracted at the given order.  Let us introduce the
following decomposition of the renormalization constants
\begin{eqnarray}
&& 
\frac{m_t(\mu^2)}{M_t} - 1 =  \delta^\alpha + \delta^{\alpha_s} +  \delta^{\alpha \alpha_s} + \cdots
\; , 
\frac{\alpha_{MS}(\mu^2)}{\alpha(M_Z)} - 1  = 
Z_e^\alpha + Z_e^{\alpha \alpha_s} + \cdots
\nonumber \\ && 
\frac{\sin^2 \theta_W^{MS}(\mu^2)}{\sin^2 \theta_W^{OS}} - 1  = 
Z_\theta^\alpha +  Z_\theta^{\alpha \alpha_s} + \cdots
\; , 
\frac{m_W^2(\mu^2)}{M_W^2} - 1  =  Z_W^\alpha + Z_W^{\alpha \alpha_s} + \cdots
\nonumber \\ && 
\overline{\Delta}r 
= \overline{\Delta}r ^\alpha + \overline{\Delta}r^{\alpha \alpha_s} + \cdots
\label{relation}
\end{eqnarray}
and shortly describe each term.  The first relation corresponds to
(5.57) in~\cite{our}.  The relation between $\overline{\rm MS}$ and
on-shell values of the electric charge, $Z_e$, includes besides the
perturbative corrections also the nonperturbative contribution from the
hadrons~\cite{EJ}.  So, in our notation, the factor $Z_e^{\alpha \alpha_s}$ 
includes only the perturbative contribution from the massive
top-quark, at the same time the factor
$Z_e^{\alpha}$ includes 
the contribution from five massless quarks, the massive leptons 
and the nonperturbative contribution (a recent numerical value is given
in~\cite{J03}).  The renormalization constant of the Weinberg angle
$\theta_W$ is related with the renormalization of the masses of the $Z$-
and $W$-bosons via $$ Z_\theta^{\alpha} +
Z_\theta^{\alpha \alpha_s} \equiv - \frac{\cos^2 \theta_W^{OS}}{\sin^2
\theta_W^{OS}}
\left[ 
Z_W^\alpha - Z_Z^\alpha + Z_W^{\alpha \alpha_s} - Z_Z^{\alpha \alpha_s}
\right], 
$$
where the perturbative contributions 
to $Z_W^{\alpha \alpha_s}, Z_Z^{\alpha \alpha_s}$
can be extracted from \cite{QCD} (see also~\cite{poleII}) and 
nonperturbative effect is given in~\cite{J}.
The $O(\alpha)$ and $O(\alpha \alpha_s)$ corrections to $\overline{\Delta}r$ 
are given in~\cite{Sirlin80} and~\cite{QCD}, respectively, and 
$\overline{\Delta} r^{\alpha \alpha_s}$ includes only 
the perturbative contribution from the quarks.

Using the decomposition (\ref{relation}) we deduce
\begin{eqnarray}
&& \hspace{-7mm}
\frac{y_t(\mu^2)}{M_t 2^{3/4}G_F^{1/2}} \!-\! 1 
\!=\! 
\delta^{\alpha} \!+\! \delta^{\alpha_s}
\!+\! \frac12 \Biggl[ Z_e^{\alpha} \!-\! Z_W^{\alpha} \!-\! Z_\theta^{\alpha} \!-\!  \overline{\Delta}r^{\alpha} \Biggr]
\nonumber \\ && \hspace{-7mm}
\!+\! \delta^{\alpha \alpha_s}
\!+\! \frac{1}{2} \delta^{\alpha_s} 
\Biggl[ Z_e^\alpha \!-\! Z_W^\alpha \!-\! Z_\theta^\alpha \!-\! \overline{\Delta}r^\alpha \Biggr]
\!+\! \frac{1}{2} \Biggl[ Z_e^{\alpha \alpha_s} \!-\! Z_W^{\alpha \alpha_s} 
                                           \!-\! Z_\theta^{\alpha \alpha_s} 
                                           \!-\!  \overline{\Delta} r^{\alpha \alpha_s} \Biggr] .
\label{yukawa}
\end{eqnarray}

The first line of the relation (\ref{yukawa})
corresponds to Eq.~(2.13) of~\cite{Yukawa}.
In the second line, the first term is our recent result 
(see Eq.~(5.57) in~\cite{our}), 
the second term is simply the product of the one-loop 
QCD correction $\delta_{\alpha_s} = 16/3 - 4 \ln \frac{m_t^2}{\mu^2}$
and the one-loop electroweak corrections and the last 
terms can be extracted from~\cite{QCD}. 
The result is relatively lengly 
and will be presented in a forthcoming publication. 

\noindent
{\bf Acknowledgments.} 
We would like to thank the organizers of the Conference 
for their very warm hospitality.
M.Yu.K. is specially grateful to Janusz~Gluza for 
his help concerning the participation at this conference. 
Furthermore, we thank I.~Ginzburg and G.~Passarino 
for interesting discussions. 
M.K.'s research was supported in part by 
RFBR grant \# 04-02-17192.


\end{document}